# Correlated photon-pair generation in reverse-proton-exchange PPLN waveguides with integrated mode demultiplexer at 10 GHz clock

Qiang Zhang[1], Xiuping Xie[1], Hiroki Takesue[2], Sae Woo Nam[3], Carsten Langrock[1], M. M. Fejer[1], Yoshihisa Yamamoto[1]

[1]*Edward L. Ginzton Laboratory, Stanford University, Stanford, California 94305*
[2]*NTT Basic Research Laboratories, NTT Corporation, 3-1 Morinosato Wakamiya, Atsugi, Kanagawa 243-0198, Japan*
[3]*National Institute of Standards and Technology, 325 Broadway, Boulder, Colorado 80305*
*qiangzh@stanford.edu*

**Abstract:** We report 10-ps correlated photon pair generation in periodically-poled reverse-proton-exchange lithium niobate waveguides with integrated mode demultiplexer at a wavelength of 1.5-μm and a clock of 10 GHz. Using superconducting single photon detectors, we observed a coincidence to accidental count ratio (CAR) as high as 4000. The developed photon-pair source may find broad application in quantum information systems as well as quantum entanglement experiments.

---

**References and links**
1. S. Tanzilli, H. De Riedmatten, W. Tittel, H. Zbinden, P. Baldi, M. De Micheli, D.B. Ostrowsky and N. Gisin, "Highly efficient photon-pair source using periodically poled lithium niobate waveguide", Electronics Letters **37**, 26-28 (2001)
2. K. Sanaka, K. Kawahara, and T. Kuga, "New High-Efficiency Source of Photon Pairs for Engineering Quantum Entanglement", Phys. Rev. Lett. **86**, 5620-5623 (2001).
3. A. Yoshizawa, R. Kaji, and H. Tsuchida, "Generation of polarization-entangled photon pairs at 1550 nm using two PPLN waveguides", Electron. Lett. **39,** 621-622 (2003).
4. H. Takesue, K. Inoue, O. Tadanaga, Y. Nishida and M. Asobe, "Generation of pulsed polarization-entangled photon pairs in a 1.55-μm band with a periodically poled lithium niobate waveguide and an orthogonal polarization delay circuit", Optics Letters **30**, 293-295 (2005)
5. T. Honjo, H. Takesue and K. Inoue, "Generation of energy-time entangled photon pairs in 1.5-um band with periodically poled lithium niobate waveguide", Opt. Express **15**, 1679 (2007).
6. A. K. Ekert, "Quantum cryptography based on Bell's theorem," Phys. Rev. Lett. **67**, 661-663 (1991).
7. C. H. Bennett, G. Brassard, N. D. Mermin, "Quantum cryptography without Bell's theorem," Phys. Rev. Lett. **68**, 557-559 (1992).
8. C. H. Bennett, G. Brassard, C. Crepeau, R. Jozsa, A. Peres, and W. Wootters, "Teleporting an Unknown Quantum State via Dual Classical and EPR Channels", Phys. Rev. Lett. **70**, 1895-1899 (1993).
9. D. Bouwmeester, J. W. Pan, K. Mattle, M. Eibl, H. Weinfurter and A. Zeilinger, "Experimental Quantum Teleportation", Nature **390**, 575-579 (1997).
10. K. Parameswaran, R. Route, J. Kurz, R. Roussev, M. Fejer and M. Fujimura, "Highly efficient second-harmonic generation in buried waveguides formed by annealed and reverse proton exchange in periodically poled lithium niobate", Optics Letters, **27**, 179-181 (2002)
11. A. J. Miller, S. W. Nam, J. M. Martinis and A. Sergienko, "Demonstration of a low-noise near-infrared photon counter with multi-photon discrimination", Appl. Phys. Lett. **83**, 791-793 (2003)
12. X. Xie and M. Fejer, "Two-spatial-mode parametric amplifier in lithium niobate waveguides with asymmetric Y junctions", Optics Letters **31**, 799-801 (2006)
13. H. Takesue and K. Inoue , "1.5-μm band quantum-correlated photon pair generation in dispersion-shifted fiber: suppression of noise photons by cooling fiber", Optics Express, **13**, 7832-7839 (2005)

14. C. Liang, K. F. Lie, M. Medic, P. Kumar, R. H. Hadfield and S. W. Nam, "Characterization of fiber-generated entangled photon pairs with superconducting single-photon detectors", Optics Express, **15**, 1322-1327 (2007).

---

**1. Introduction**

A 1.5-μm correlated photon-pair source can play an important role in long distance quantum communication over optical fiber due to the low transmission loss (<0.2 dB/km) and small pulse broadening in dispersion shifted fiber (DSF). The conventional method to generate correlated photon-pairs is the use of parametric down conversion in bulk nonlinear crystals. Periodically poled lithium niobate (PPLN) waveguides have been used recently to generate correlated photon-pairs due to their high conversion efficiency compared to their bulk counterparts [1-5]. Furthermore, waveguide devices can be fiber pigtailed to achieve higher collection efficiency, as well as easy and stable operation. These correlated photon-pairs can be easily converted to entangled photon-pairs [1,3-5] for quantum cryptography [6,7] and quantum teleportation [8,9].

However, the repetition rate of previous experiments was limited to approximately 100 MHz. [2,3] Here, signal and idler photons were generated at degenerate wavelengths and in the same spatial and polarization mode. Thus, to separate them may introduce an additional 3 dB loss.

Here we present 1.5-μm correlated photon pair generation using 10-ps pump pulses at a repetition rate of 10 GHz via the parametric down conversion process in a low-loss (<0.1 dB/cm) reverse-proton-exchange (RPE) PPLN waveguide device [10]. Superconducting single photon detectors (SSPD) [11] with a small timing jitter (FWHM=65 ps) were used to detect the correlated photon pairs in the 1.5-μm-band. A monolithically integrated mode de-multiplexer using asymmetric Y-junctions is used to separate the correlated photon-pairs. [12] In our experiment, we measure the coincidence to accidental count ratio (CAR) to characterize the correlated photon-pairs. The highest CAR of our setup observed was 4000, which is about two orders of magnitude higher than was observed in previous experiments. [13,14] Furthermore, if we only consider the photon pair source itself and assume 100% quantum-efficiency single photon detectors, our setup would generate 100 MHz photon pair with a CAR of 9.5, which is the brightest photon pairs source to our knowledge.

2. Background

*2.1 Waveguide-based parametric down conversion*

Since protonated $LiNbO_3$ waveguides only support TM-polarized waves in z-cut substrates, perfect separation of the signal and idler waves via polarization de-multiplexing is not possible in near-degenerate parametric down conversion. Mode de-multiplexing using asymmetric Y-junctions, based on the adiabatic variation of the refractive index distribution along the device, can be utilized to solve this problem. Figure 1 shows the design of such a device and illustrates its functionality. The pump wave in the $TM_{00}$ mode is coupled into the narrow input arm of the Y-junction and is adiabatically converted into $TM_{10}$ mode of the main waveguide. The pump then traverses the quasi-phase-matching grating and generates photon-pairs in different spatial modes (signal in $TM_{10}$ and idler in $TM_{00}$, respectively). The idler photon in the $TM_{00}$ mode will exit via the wide arm of the Y-junction's output port, while the signal photon in the $TM_{10}$ mode will be coupled into the narrow arm. Both output modes are then adiabaticaly transformed to $TM_{00}$ modes with identical sizes. For wavelengths near 1.5-μm, an extinction ratio of >30 dB between the two output modes can be achieved with current fabrication technologies. [12]

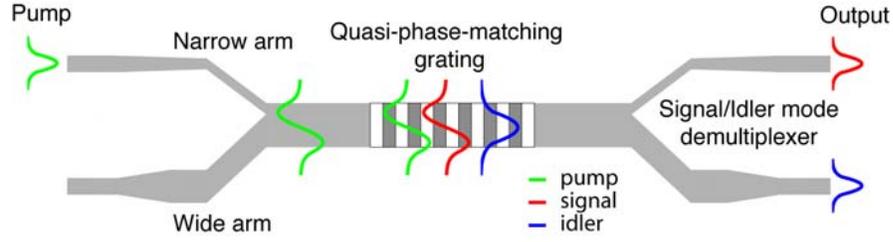

Fig. 1. Asymmetric Y-junction device for parametric down conversion.

*2.2 Time correlation measurement and CAR*

Following Reference [13,14], we measured the CAR to characterize our correlated photon-pair source. If the pump pulse is transform limited, the correlated photon-pair must naturally be generated in a single temporal mode set by the pump pulse duration and the photon-pair number distribution obeys a Poissonian distribution. Figure 2a is the histogram of the signal photon's detection time, which demonstrates that the timing jitter of the SSPD detector (60 ps) is smaller than the pulse interval (100ps), while Figure 2b is the histogram of the photon-pair coincidence's detection time. The highest peak in Figure 2b results from the coincidence caused by photon-pairs generated by the same pump pulse. In fact, it also includes the accidental count caused by multi-photon-pair generation or noise photons in the same pulse. Other small peaks in different time slots are accidental counts caused by photon-pairs in different pulses and dark counts. The CAR is simply the ratio between the counts in the highest peak and in any other small peak.

Suppose the average number of photon-pairs per pulse is $\mu$ and the laser repetition frequency is $\nu$. The average noise photon numbers per pulse are $\mu_s$ and $\mu_i$ for the signal and idler channel, respectively, which stems mainly from the residual pump light and various nonlinear processes other than down conversion. The average number of dark counts per pulse in the two channels are $t*d_s$ and $t*d_i$, respectively, where $t$ is the time window set by the timing jitter of the SSPD, while $d_s$ and $d_i$ are the dark count rates of the two SSPDs. The overall single-photon detection efficiencies, (including collection and detection quantum efficiency), are $\eta_s$ and $\eta_i$. The coincidence count rate due to the photon-pair can then be expressed as:

$$C = \nu * \mu * \eta_s * \eta_i \qquad (1).$$

The accidental count rate is:

$$C_a = \nu * ((\mu + \mu_s)*\eta_s + t*d_s)*((\mu + \mu_i)*\eta_i + t*d_i) \qquad (2).$$

Therefore the coincidence in the highest peak to accidental count ratio is:

$$CAR = (C + C_a)/C_a = \frac{\mu * \eta_s * \eta_i}{[(\mu+\mu_s)*\eta_s + t*d_s]*[(\mu+\mu_i)*\eta_i + t*d_i]} + 1 \qquad (3).$$

From Eq. (3), we can see that the CAR mainly depends on $\mu_{s/i}$ and $t*d_{s/i}$ when the collection and detection efficiencies are being held fixed. When the pump power is high, $\mu$ is large and the noise is mainly due to the multi-photon-pair generation processes and the residual pump light, i.e. $\mu_{s/i}$. When the pump power is low, the noise is dominated by the detector dark count rate per time window, i.e. $t*d_{s/i}$.

As mentioned in Ref. [13], the CAR can provide an estimate of the performance of a photon-pair source in a quantum communication system. For example, it can be used to estimate the error rate of quantum key distribution based on quantum entanglement. [13]

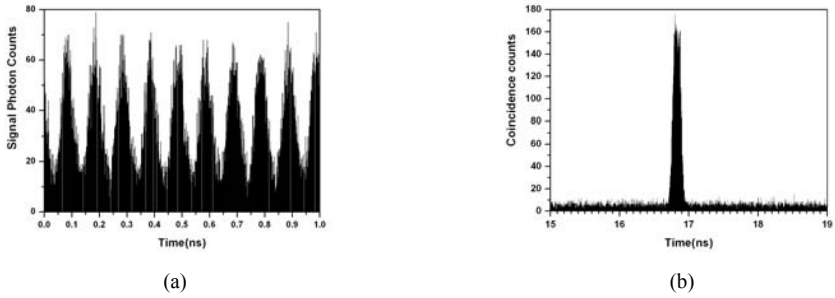

(a)                                       (b)

Fig. 2. Histograms of the signal photon (a) and coincidence of the correlated photon pair (b) in the time domain measured by an Ortec 9308 time interval analyzer (TIA). In (a), the 0.1-ns period shows the laser's repetition rate and the 60-ps FWHM of the histogram represents the timing jitter of the SSPD. In (b), the FWHM of the histogram of photon pair is about 0.1 ns, which is not only due to the timing jitters of the two SSPDs for the photon pair detection, but also the waveguide dispersion.

## 3. Experiment

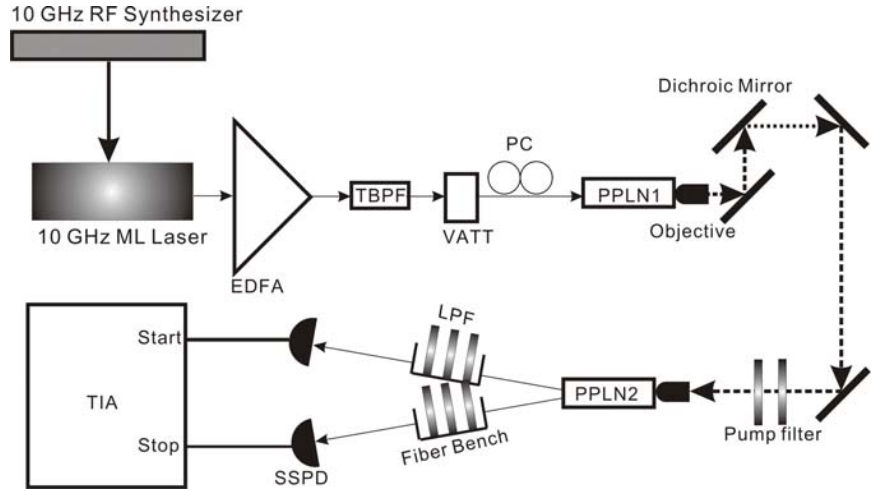

Fig. 3. Diagram of the experimental setup. TBPF: tunable band-pass filter. VATT: variable fiber attenuator. PPLN1: a periodically-poled lithium niobate waveguide for second harmonic generation of the pump source. Pump filter: 780 nm bandpass filter to remove 1.5-μm background. PPLN2: a fiber pigtailed asymmetric Y- junction periodically-poled lithium niobate waveguide for parametric down-conversion. LPF: long-pass filter to remove the residual pump light. SSPD: superconducting single-photon detector. TIA: time interval analyzer. Solid lines represent fiber and dotted lines represent free space propagation.

Figure 3 shows our experimental setup. An actively mode-locked fiber laser (Calmar Optcom PSL-10FTSTF11) triggered by a 10-GHz RF synthesizer (HP 8371B) was used to provide a 10-GHz pump pulse train. After being amplified by an erbium-doped fiber amplifier (EDFA), the laser pulses with a 10-ps FWHM duration and 1563-nm center wavelength pass through a tunable bandpass filter to remove the background noise from the EDFA. The pump is then frequency doubled in the first PPLN waveguide chip. Since these waveguide devices only accept TM-polarized light, an in-line fiber polarization controller is used to adjust the polarization of the input. The residual pump is attenuated by 180 dB using dichroic mirrors and pump filters. The second harmonic (SH) wave is then launched into the second PPLN waveguide as the pump pulse for the parametric down conversion process. This second PPLN waveguide has an asymmetric Y-junction at both input and output for mode launching and mode de-multiplexing as shown in Fig. 1. The two output ports are fiber pigtailed to improve the collection efficiency and stability of operation. A fiber bench with several long-pass filters is inserted in each output fiber of the second PPLN chip to eliminate residual pump light (the second harmonic wave). The down-converted photons are then detected by two SSPDs and analyzed by a time interval analyzer (TIA).

The two PPLN chips are temperature controlled to satisfy the phase matching conditions. The first chip's temperature was set to 90 °C and the second to 130 °C.

The SSPDs used in this experiment consist of a 100-nm wide, 4-nm thick NbN superconducting wire, which is coupled to a 9-μm core single-mode fiber.[11] The packaged detector is housed in a closed-cycle cryogen-free refrigerator with an operating temperature of 3 K. The quantum efficiency and dark count rate of the SSPDs depend on an adjustable bias current. In the experiment, we set the bias current to reach a quantum efficiency of 1.1% and 3.8%, respectively, and a dark count rate of less than 100Hz. These SSPDs have an inherently small Gaussian timing jitter of 65 ps FWHM that agrees well with the histograms shown in Figure 2.

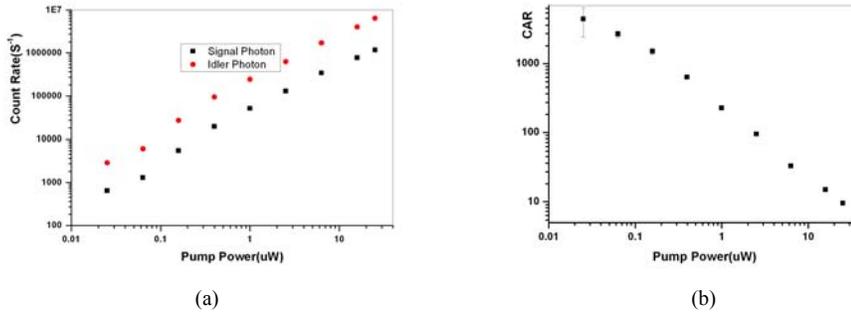

(a)  (b)

Fig. 4. (a) Single photon count rates from two superconducting single photon detectors respectively detecting the two photons in a correlated pair. Black squares represent signal photon and red circles idler photons. The difference in count rates at any given pump power is mainly due to different quantum efficiency of the two SSPDs in our experiment. Fig. 4. (b) CAR of the correlated photon pair generation. The x-axis is the 781.5nm pump power generated by the first SH chip. All data points are taken with a 5E8 start signal pass except for the one with highest CAR, where we only took 5E7 signal instead due to the low count rate. This is also the main reason for its large error bar.

Figure 4(a) shows the count rate of the two detectors as a function of the pump power. An Ortec 994 dual counter is used in the experiment to count the photon number in both signal and idler channels. When the input power of the first PPLN waveguide (SH chip) was set to 25 mW, the 780 nm pump power coupled into the second PPLN waveguide (Y-junction chip) was about 25 μW, resulting in a signal photon count rate of 1.1 MHz and an idler photon count rate of 4.8 MHz. The total passive loss (including absorption, reflection and scattering loss) in the Y-junctions chip was about 3 dB which is similar to a conventional straight waveguide of the same length. The collection efficiency including the fiber pigtailing insertion loss and the transmission loss from the long pass filter and fiber bench was approximately 6 dB. The coincidence rate of the photon pairs at this pump power was 10 kHz measured by the TIA in Figure 3.

Figure 4(b) shows the measured CAR of the correlated photon-pairs versus pump power. The electronic signal from the SSPD for the idler photon is taken as the start signal to the TIA and the signal of the SSPD for the signal photon served as the stop signal. An Ortec DB463 delay generator provides a 50-ns delay for the stop signal in order to match the 50 ns interpolator dead time of the TIA start signal. The TIA can show the start and stop signal's coincidence in the time domain. As explained in Section 2.2, the CAR can be measured by comparing the peak coincidence counts to another time slot's accidental counts within the same time window. We chose a 60-ps time window similar to the SSPD's timing jitter. Due to the 0.1 ns FWHM of the coincidence histogram, we chose a 0.2 ns time slot. Furthermore, in order to reduce statistical errors, we took the average of 10 peak accidental counts in 10 different time intervals as the final accidental counts.

With 25 μW coupled pump power, a photon-pair rate of 0.12 pair per pulse and a CAR of 9.5 was measured. The highest CAR we obtained was 4452, corresponding to a pump power of 0.02 μW. The average number of photon-pairs per pulse in this case was 0.0002. With the low pump power, the dark count rate per pulse dominated the accidental count rate. This CAR is about two orders of magnitudes higher than in previously demonstrated systems. [13,14] The reason for this improvement is the high repetition rate (10 GHz), narrow time window (60 ps) and low dark count rate of the SSPD (100 Hz). These provide a much smaller $t*d_{s/i}$ in Eq. 3.

## 4. Conclusion

In summary, for the first time we used pump pulses at a repetition rate of 10 GHz to generate 1.5-μm-band correlated photon-pairs using integrated mode de-multiplexer in RPE PPLN waveguides. Using SSPDs and waveguides with asymmetric Y-junctions, we obtained a high CAR of 4452 for the generated photon pairs. These results may allow us to generate entangled photon pairs with very high yield and visibility. It will find immediate application in long distance fiber-based quantum communication and various quantum optics experiments that require a high CAR.

**Acknowledgement**

The authors thank M. J. Kim and N. Y. Kim for lending the 10 GHz RF synthesizer. This research was supported by the U.S. Air Force Office of Scientific Research through contracts F49620-02-1-0240, the MURI center for photonic quantum information systems (ARO/ARDA program DAAD19-03-1-0199), the Disruptive Technology Office (DTO), SORST, Science and Technology Agency of Japan (JST) and the NIST quantum information science initiative. We acknowledge the support of Crystal Technology, Inc.